\newif\ifsubmit
\submitfalse

\newif\ifdraft
\drafttrue

\ifdraft
\documentclass[draft]{llncs}
\else
\documentclass{llncs}
\fi

\usepackage[T1]{fontenc}
\usepackage[utf8]{inputenc}
\usepackage{lmodern}
\usepackage[scaled=.8]{beramono}
\usepackage{latexsym}
\usepackage{textcomp}
\usepackage{amssymb}
\usepackage[final]{graphicx}
\usepackage[final]{listings}
\usepackage{lstsemantic}
\usepackage{tabularx}

\usepackage[british]{babel}
\usepackage[inline]{enumitem}
\usepackage[babel]{csquotes}

\usepackage[obeyDraft,textsize=tiny]{todonotes}

\usepackage[backend=bibtex,hyperref=auto,
firstinits=true,style=numeric,
urldate=iso8601,isbn=false,doi=false]{biblatex}
\DeclareNameAlias{author}{last-first}
\DeclareNameAlias{editor}{last-first}
\DeclareNameAlias{translator}{last-first}
\DeclareFieldFormat{labelnumberwidth}{#1.}
\DeclareFieldFormat{title}{#1}
\DeclareFieldFormat
  [article,inbook,incollection,inproceedings,patent,thesis,unpublished]
  {title}{#1}
\DeclareFieldFormat{journaltitle}{#1}
\newbibmacro*{volume+number+eid}{%
  \printfield{volume}%
  \iffieldundef{number}
    {}
    {(%
      \printfield{number}%
     )%
    }%
  \setunit{\addcomma\space}%
  \printfield{eid}}
\DeclareFieldFormat{url}{\url{#1}}

\usepackage{hyperref}

\begin{document}

\title{The WDAqua ITN – Answering Questions using Web Data\thanks{The authors of this abstract are the coordinators of the project.
Representatives of further project partners will also be present at ESWC.}}
\author{Christoph Lange\inst{1}
\and Saeedeh Shekarpour\inst{1}
\and Sören Auer\inst{1}}
\institute{%
University of Bonn, Germany
\email{\{langec,shekarpour,auer\}@cs.uni-bonn.de}
}

\maketitle

\section{Motivation and Objectives}

WDAqua is a Marie Skłodowska Curie Innovative Training Network (ITN) that employs 15 PhD students overall.

The H2020 work programme aims at increasing citizens' participation in the digital society and making infrastructures “smart”.
These processes are increasingly data-driven.
Our central motivation is that sharing, connecting, managing, analysing and understanding data on the Web will enable better services for citizens, communities and industry.
However, turning web data into successful services for the public and private sector requires skilled web and data scientists, and it still requires further research.
WDAqua aims at advancing the state of the art by intertwining training, research and innovation efforts, centered around one service: data-driven question answering.
Question answering is immediately useful to a wide audience of end users, and we will demonstrate this in settings including e-commerce, public sector information, publishing and smart cities.
Question answering also covers web science and data science broadly, leading to transferrable research results and to transferrable skills of the researchers who have finished our training programme.

Simplified, steps to answering a question are: 1) understanding a spoken question, 2) analysing the question's text, 3) finding data to answer the question, and 4) presenting the answer(s).
To ensure that our research improves question answering overall, every individual research project connects at least two of these steps.
Intersectoral secondments (within a consortium covering academia, research institutes and industrial research; see below) as well as network-wide workshops, R\&D challenges and innovation projects further balance ground-breaking research and the needs of society and industry.
Training-wise these offers equip early stage researchers with the expertise and transferable technical and non-technical skills that will allow them to pursue a successful career as an academic, decision maker, practitioner or entrepreneur.

\section{Facts}

WDAqua is funded under EU grant number 642795 and runs from January 2015 to December 2018.\footnote{See \url{http://wdaqua.informatik.uni-bonn.de}}
It involves the following 6 partners, ordered roughly according to the steps of question answering outlined above: 
Fraunhofer Institute for Intelligent Analysis and Information Systems (IAIS), Germany (pluggable question answering architecture, understanding spoken questions);
University of Bonn, Germany (coordinator; finding high-quality data, analysing natural language questions);
National and Kapodistrian University of Athens, Greece (handling context and question ambiguity);
Université Jean Monnet Saint-Étienne, France (retrieving and linking dynamic data);
University of Southampton, UK (explaining and visualising answers, interactive dialogs); and
the Open Data Institute (ODI), UK (further exploitation of answers; e.g., licensing or pricing).

Overall, they will employ 15 PhD students, whose recruitment has almost been completed at the time of this writing (April 2015).
Each PhD student will spend secondments of several months with other network partners, including six industrial partner organisations from different application domains:
Wolters Kluwer Deutschland, Germany (legal publishing);
Unister, Germany (e-commerce);
Data Publica, France (government and enterprise data);
Antidot, France (information retrieval);
Fundacio Barcelona Media, Spain (hosting the European Yahoo research labs); and
Athens Technology Center, Greece (web engineering).
 
\section{Networking Potential}

Outreach to the scientific community, to industry and to society is an essential factor in every ITN.
In the current phase, where the actual research has not yet started (most PhD students will start around July 2015), the consortium is obviously most interested in attracting further \textit{input} from the community; whereas in later phases we will be able to contribute results to the community and to the general public.
Ways of feeding input into the network include:

\noindent\textbf{Teaching in training events} hosted by the project.
  We will have one learning week plus one research and development week for all PhD students every year, as well as research workshops.
  We generally aim at addressing a larger audience than just 15 PhD students.
  We will achieve this by co-locating our learning weeks with summer schools open to the general public; e.g., our first learning week will be co-located with the 2015 Web Intelligence Summer School on Web Science \& Question Answering with the Web\footnote{\url{http://wiss.univ-st-etienne.fr/}}.\\
\noindent\textbf{Providing systems and data} to be reused in the WDAqua PhD projects.
  We have a broad understanding of question answering and consider the whole pipeline from data provision to communicating answers.
  Bootstrapping this pipeline will benefit from the ability to reuse existing components, e.g., for natural language processing.\\
\noindent\textbf{Hosting PhD students} for internships.
  We are aware of the broad applicability of question answering, and thus open to suggestions from application domains complementing those represented within the network.\\
\noindent Being a \textbf{member of the advisory board}.
  We have several high-profile international members already but are still open to applications.

\printbibliography
\end{document}

